\documentclass[%
reprint,
 twocolumn
 nofootinbib,
 nobibnotes,
 amsmath,amssymb,
 aps,
 prl,
 floatfix,
 tightenlines,
 10pt
]{revtex4-1}

\bibpunct{[}{]}{;}{n}{}{}
\usepackage{graphicx}
\usepackage{dcolumn}
\usepackage{bm}
\usepackage[hidelinks]{hyperref}
\usepackage{physics}
\usepackage{float}
\usepackage{bbold}
\usepackage{nicefrac}
\usepackage{soul}
\usepackage{placeins}
\usepackage[normalem]{ulem}
\usepackage{xcolor}
\usepackage{url}

\definecolor{ABcolor}{cmyk}{1,0,0,0.6}

\definecolor{green1}{rgb}{0,0.6,0}

\definecolor{AMColor}{rgb}{0.16,0.23,0.55}

\definecolor{mColor}{rgb}{1.0,0,0}
\newcommand{\modified}[1]{{#1}}

\begin{document}
\title{Vortex nucleation barrier in superconductors beyond the Bean-Livingston approximation: A numerical approach for the sphaleron problem in a gauge theory}

\author{Andrea Benfenati} \email{alben@kth.se}
\affiliation{Department of Theoretical Physics, The Royal Institute of Technology, Stockholm SE-10691, Sweden}

\author{Andrea Maiani} \email{andrea.maiani@nbi.ku.dk}
\affiliation{Department of Theoretical Physics, The Royal Institute of Technology, Stockholm SE-10691, Sweden}

\author{Filipp N. Rybakov} 
\affiliation{Department of Theoretical Physics, The Royal Institute of Technology, Stockholm SE-10691, Sweden}

\author {Egor Babaev}
\affiliation{Department of Theoretical Physics, The Royal Institute of Technology, Stockholm SE-10691, Sweden}

\date{\today}

\begin{abstract}
The knowledge of vortex nucleation barriers is crucial for applications of superconductors, such as single-photon detectors and superconductor-based qubits.
Contrarily to the problem of finding energy minima and critical fields, there are no controllable methods to explore the energy landscape, identify saddle points, and compute associated barriers. Similar problems exist in high-energy physics where the saddle-point configurations are called sphalerons.
Here, we present a generalization of the string method to gauge field theories, which allows the calculation of energy barriers in superconductors. We solve the problem of vortex nucleation, assessing the effects of the nonlinearity of the model, complicated geometry, surface roughness, and pinning.
\end{abstract}
\maketitle
%
%
In type-II superconductors, the Meissner state, characterized by the total magnetic field expulsion in the bulk, is stable up to the lower critical field $H_\mathrm{c1}$. Above it, quantum vortices appear.
However, the Meissner state can survive as a metastable state, causing the phenomenon known as magnetic superheating~\cite{schmidt2013physics}. The presence of fluctuations can trigger the spontaneous decay of the metastable Meissner state through vortex formation. 
This phenomenon is the effect of a surface barrier, which hinders vortex nucleation from the sample boundaries~\cite{bean1964surface,schmidt2013physics}. The barrier disappears when the applied magnetic field exceeds a critical value, the nucleation field. 
\modified{In contrast to the potential barrier, the nucleation field is not only amenable to analytical treatment~\cite{de1965vortex, kramer1968stability, Fink1969, Christiansen1969,  kramer1973breakdown, Dolgert1996, catelani2008temperature,  transtrum2011superheating, liarte2017theoretical}, but can be calculated with a controlled numerical accuracy from iterative simulations~\cite{PhysRevB.52.R15719, PhysRevB.76.024509, Pack2020}. }

The full problem of the vortex entry barrier is not solved even for the simplest case of a semi-infinite superconductor with an ideal surface.
The Bean-Livingston estimate \cite{bean1964surface} relies on the London model, but the result depends on the choice of cutoff, as emphasized by the authors themselves. 
The Bean-Livingston approach has been extended to other geometries  \cite{galaiko1966formation, Fetter1980}.The effect of surface roughness is considerably more complicated. Only some approximate analytical approaches within the London model were advocated in Ref. \cite{bass1996effect}. However, roughness alters the core structure and therefore a controllable method to solve the full nonlinear model is needed.

Fundamentally, the problem of a vortex entry barrier consists of finding the \emph{sphaleron}, i.e., the saddle point which separates two stable states in a gauge theory \cite{rubakov2009classical, manton2004topological, Nastase2019,Kolb1994,shifman_2012}.
In the case of vortex nucleation, it is an energy maximum of the minimum energy path between two states with different phase windings.

\modified{The knowledge of potential barriers for vortex nucleation is crucial for superconductor applications. The applications include current transmission, where the dissipationless state is lost if free vortices form, and  superconductive rf cavities for particle accelerators where superconductive magnets often operate in a superheated state \cite{liarte2017theoretical}.}
Recently, the problems of vortex entry barriers appeared in quantum technologies.
In superconducting single-photon detectors, it is believed that the principle of operation consists in the creation of a current-carrying state with a small potential barrier for vortex entry so that a single photon creates a vortex, hence a  detectable signal~\cite{natarajan2012superconducting, doi:10.1063/1.5000001}.
However, small barriers yield spontaneous vortex nucleation caused by fluctuations, resulting in dark counts. The ability to calculate potential barriers would allow designing devices with significantly improved performances. 
Likewise, the knowledge of vortex entry barriers is crucial to design superconducting topological qubits~\cite{Kitaev_2001, RevModPhys.80.1083, PhysRevLett.122.187702, Lutchyn2010, Lutchyn2018}, where the geometry of the device plays a central role.
\modified{
These devices often operate at temperatures near absolute zero. However, while a microscopically derived Ginzburg-Landau model applies only close to the critical temperature, many aspects of low-temperature vortex physics may under certain conditions be fittable by effective Ginzburg-Landau-type models \cite{silaev2012microscopic}. }
In this Rapid Communication, we generalize to gauge theories the simplified string method \cite{weinan2007simplified}.
This allows us to perform surface barrier calculations in type-II superconductors for vortex nucleation ($\Delta F_\textrm{n}$) and escape ($\Delta F_\textrm{e}$), by computing the minimum energy path of the transition in the Ginzburg-Landau theory.
The results differ from the Bean-Livingston theory \cite{bean1964surface}, which neglects the vortex-core and nonlinear effects. More importantly, the method allows taking into account the effect of surface roughness and the presence of pinning in the computation of the vortex nucleation barrier and the superheating field.
%
%
The Ginzburg-Landau free-energy functional describing the superconductor in dimensionless units is
\begin{equation}
\label{eq:SI-GLmodel}
\begin{split}
f[\vb a, \psi] = \int_\Omega \dd{\Omega} \left( \frac{1}{2} \left| \left(-i \frac{2}{\kappa} \nabla + \vb a \right) \psi \right|^2\right. \\ \left. + \frac{1}{2}\left( 1 - \abs{\psi}^2\right)^2 + \frac{(\curl \vb a - \vb h)^2}{2}\right)\,.
\end{split}
\end{equation}
The complex field $\psi = \abs{\psi}e^{i\theta}$ describes the state of the superconductor. The vector potential $\vb a$ is related to the magnetic field by $\vb b = \curl \vb a$. The coefficient $\kappa = \frac{\lambda}{\xi}$ is the Ginzburg-Landau parameter, i.e., the ratio of  the magnetic field penetration depth $\lambda$ and the coherence length $\xi$ (we use the definition of coherence length  with a factor $\sqrt{2}$ absorbed as in Ref.\cite{svistunov2015superfluid}). The external magnetic field $\vb H$ is expressed  in units of the thermodynamic critical field $H_\mathrm{c}$, i.e., $h = H/H_\mathrm{c}$.
The free energy $f = F/F_0$  is expressed in units of $F_0$, which in SI units is $F_0 =  \mu_0 H_\textrm{c}^2 \lambda^2 d$, where
$d$ represents the thickness of the sample and $\mu_0$ is the vacuum permeability. Quantities in capital letters are intended in SI units, while variables in lower case are dimensionless.

In a  finite system, the Meissner state can survive in a meta-stable way for higher fields than $H_\textrm{c1}$, i.e., up to the \emph{spontaneous nucleation field} $H_\textrm{n}$. In the absence of fluctuations, only when this field is exceeded, vortices nucleate from the boundaries. To introduce a vortex in the system, when $H<H_\textrm{n}$, we need to overcome an energy barrier due to the surface.
%
%
The calculation of this barrier is a nonlinear problem that is not amenable to analytical treatment. In this work, we generalize a numerical approach, which was originally developed within the molecular dynamics field to study transitions between two metastable states through the identification of a \emph{minimum energy path} \cite{vanden2010transition}. 
The path is considered in the configuration space of the system, and it is parametrized by the \emph{transition coordinate} $s \in [0,1]$.
If one denotes by $\vb q(s) = [\vb a(s,\mathbf{r}), \psi(s,\mathbf{r})]$, the state of the system, then $\vb q(0)$ and $\vb q(1)$ are two equilibrium solutions corresponding to the minima of the  Hamiltonian in Eq. \eqref{eq:SI-GLmodel}.
By varying $s$ from 0 to 1, we observe the transition of the system from the initial state to the final state.
One can assume that a potential force 
$\mathbf{g} \equiv -\grad f = -\Big(\frac{\delta f}{\delta{\vb a}}, \frac{\delta f}{\delta{\psi^*}}\Big)$ 
acts on each point of the curve $\vb q(s)$.
The minimum energy path is a trajectory in which the force $\vb{g}$ acting on each point is uniquely directed along the tangent vector  $\pdv*{\vb q}{s}$, i.e., 
\begin{equation}
\abs{\grad f \cdot \pdv{\vb q}{s}} = \norm{\grad f}\norm{ \pdv{\vb q}{s}}\,, \qquad \forall s \in [0,1]\,.
\label{eq:mepRequirement}
\end{equation}
We emphasize that the optimal path defined by Eq. \eqref{eq:mepRequirement} does not correspond to the real time dynamics. It describes the most energetically favorable transformation undertaken by the system for the transition between the initial and the final state.

To identify the minimum energy path, we started from the well-known \emph{simplified string method} \cite{weinan2007simplified} and
developed a variant applicable to gauge field theories \footnote{See Supplemental Material at [URL will be inserted by publisher] for details of the gauged string method developed for this problem, which includes Refs. \cite{BESSARAB2015335, doi:10.1063/1.1329672, koslover2007comparison, sheppard2008optimization, kramers1940brownian, langer1969statistical}}.
This algorithm evolves an initial guessed path in the configuration space, towards the minimal energy one. Consider a situation where we start from a Meissner state and end in the one-vortex state.
To construct the initial guess, we used an ansatz for the single winded vortex state.
In the initial state,  for $s=0$, the vortex is outside of the domain and in the final state, for $s=1$, it lies in the origin. 
To each value of the transition coordinate $s$, there corresponds a particular configuration of the system.
Hence, $s$ is not equivalent to the position of the center of the vortex because static vortex-core deformations correspond in general to different values of $s$. This method allows us to solve the full nonlinear problem in contrast to previous approaches, and obtain exact and quantitatively valid results.
A previous study of the vortex entry barrier, based on the London model, was carried out by Bean-Livingston \cite{bean1964surface}. However, this introduced the uncontrollable approximation of considering the vortex core as a rigid cylinder of radius $\xi$, which neglects the physics of the core and the nonlinear effects.

The energy dependence $F(s)$ is a function with one or more maxima. At the summit of the potential barrier, the configuration of the system is a saddle point of the free-energy functional of Eq .\eqref{eq:SI-GLmodel}, which in the context of gauge theories is called a \emph{sphaleron}.

Once the minimum energy path is computed, we can define the nucleation barriers $\Delta F_\textrm{n} = F_{\textrm{sphaleron}} - F_{\textrm{Meissner}}$ for a given magnitude of the external magnetic field $H$. Consequently, the nucleation field  $H_\textrm{n}$ is the external field needed to nullify the nucleation barrier, i.e., $\Delta F_\textrm{n}(H_\textrm{n}) = 0$ \footnote{
\modified{Note that the nucleation field $H_\textrm{n}$ is defined as the external field needed for suppressing the barrier and cause a single spontaneous vortex entry.
The superheating field $H_{sh}$, often referred to in literature, is defined as the upper metastability limit of the Meissner phase for large and homogeneous superconductors.
}}.
Analogous definitions can be used for the escape barrier, i.e., the energy needed to expel one vortex from the superconductor. A quantitative description of nucleation and escape barriers is given in the Appendix.
The minimum energy path contains more information than the height of the energy barrier as those paths are most likely to be followed in the nucleation process. This path helps us understand in detail how transitions between metastable states occur. 
\begin{figure}[!hbt]
    \centering
    \includegraphics[width=0.98\columnwidth]{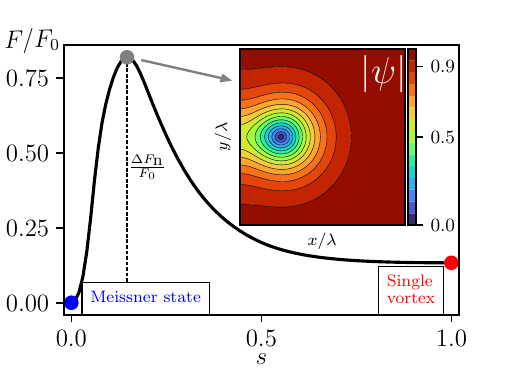} 
    \caption{
    The energy dependence and the saddle-point (sphaleron) configuration for the process of vortex entry from a straight edge in a two-dimensional superconductor.  At the peak of the energy barrier, we have a substantial vortex-core deformation which cannot be described in the London limit, where $\psi$ is assumed to be constant. This shows how this kind of process, in general, cannot be treated in the London model, i.e., within the Bean-Livingston approach. 
The variable $s$ parametrizes the minimum energy path.
In this example $H=0.6 H_\mathrm{c}$ and $\kappa=2$.
The inset shows $\abs{\psi}$ in the region when the nucleation takes place.
}
\label{fig:example_mfep_GL}
\end{figure}
In our framework, we can calculate the energy barrier free from any approximations, except fully controlled numerical errors.
%
We begin by considering a vortex entry in a 2D superconductor
with flat surfaces. We find that the vortex always enters, following the minimum energy path, from the sides of the superconductor and never from the corners.
Figure \ref{fig:example_mfep_GL} shows the free-energy profile of the process. The barrier presents a single maximum corresponding to the sphaleron. The substantial core deformation confirms that this kind of problem is not in general treatable with controlled accuracy in the London limit.

For complex materials, the surface
can have different roughness, doping and oxidation, which strongly affects the vortex entry process.
The knowledge of how impurities influence vortex physics is crucial for applications \cite{FossheimKristian}.

\begin{figure}[!hbt]
    \centering
    \includegraphics[width=0.87\columnwidth]{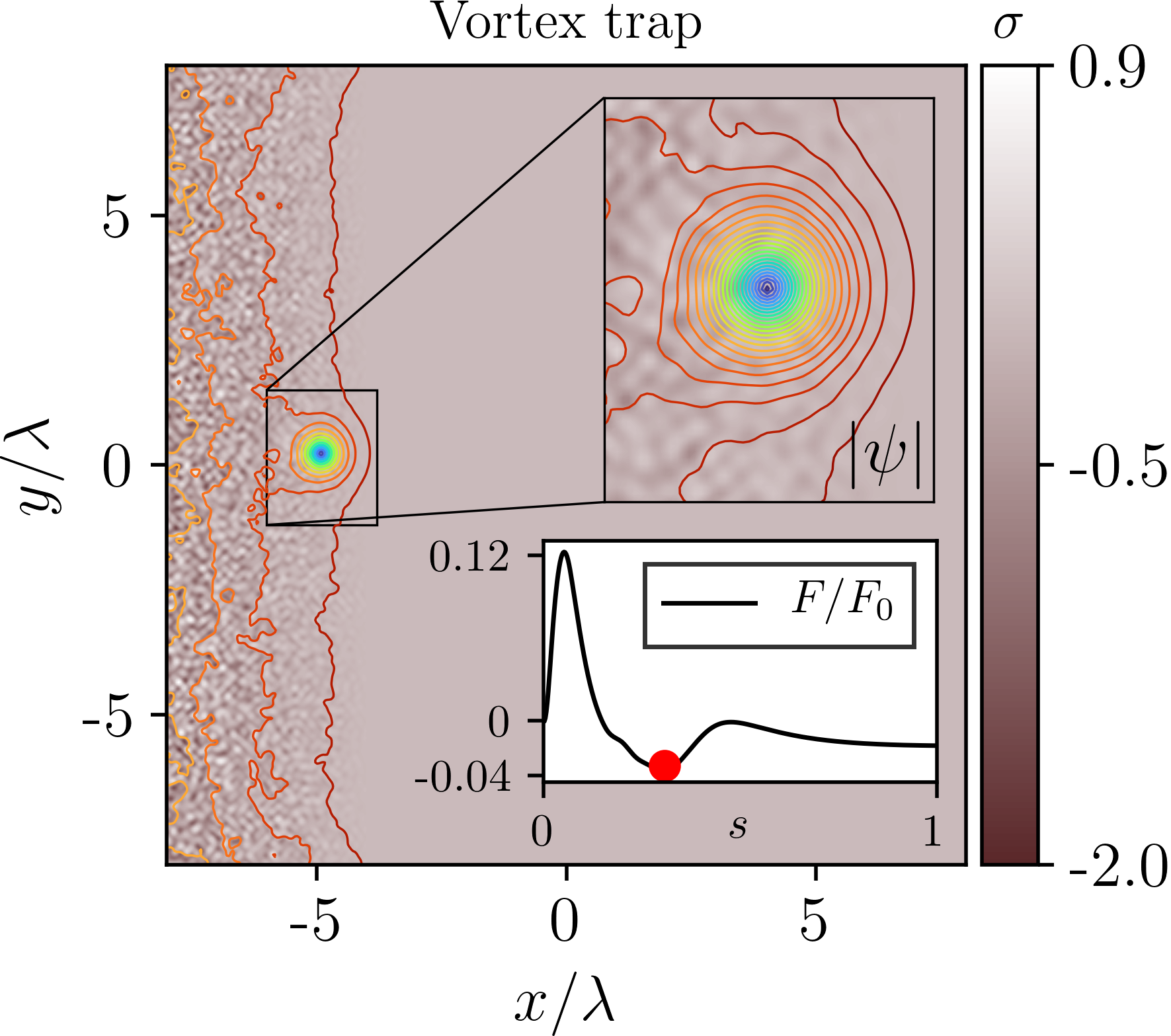} 
    \caption{Trapped vortex and corresponding energy profile for the minimum energy path in the case of a superconductor with \modified{inhomogeneities} concentrated near the edge. The impurity layout $\sigma$ is normally distributed, and it is decreasing to zero as we proceed towards the bulk. On top of it, we plot isolines of the order parameter modulus $\abs{\psi}$.
    For the color notation of isolines, see the color bar in Fig. \ref{fig:example_mfep_GL}.
    Moreover, we have $H=0.4H_\mathrm{c}$ and $\kappa=5$. 
    }
    \label{fig:impurities}
\end{figure}

\begin{figure}[!hbt]
    \centering
    \includegraphics[width=0.80\columnwidth]{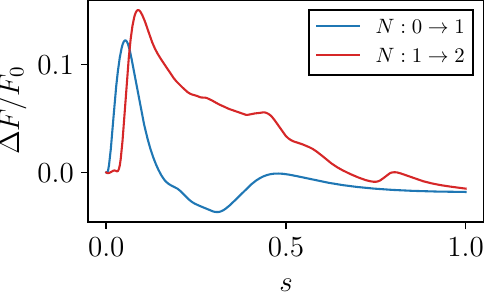}
    \caption{
    Minimum free-energy paths for the first (blue) and second (red) vortex entry in the presence of the impurity modulation of Fig. \ref{fig:impurities}. For the second entrance simulation, the initial condition was built by placing a vortex in the trap as shown in Fig. \ref{fig:impurities} (initial winding number $N=1$). Impurities pin the first vortex near the boundary, increasing the nucleation barrier of the second one. }
    \label{fig:nucleationBarriers}
\end{figure}

Let us consider randomly distributed \modified{inhomogeneities}, with a decreasing density as we enter the sample, as shown in Fig. \ref{fig:impurities}. To model \modified{inhomogeneities} in a superconductor, we follow a procedure similar to the one outlined in Refs. \cite{chapman1995ginzburg, Du1995, deang1997vortices}, where we modify the quadratic term in Eq. \eqref{eq:SI-GLmodel} accordingly,

\begin{equation}
    \frac{1}{2}(1-\abs{\psi(\vb{r})}^2)^2 \, \rightarrow \, \frac{1}{2}(1 + \sigma(\vb{r})-\abs{\psi(\vb{r})}^2)^2 \,,
\end{equation}
where $\sigma(\vb{r}) $ is the inhomogeneity distribution. We tested different models for the inhomogeneity distribution with similar results.
In the free-energy profile, depicted in Fig. \ref{fig:impurities}, there is a global minimum in between the saddle points. 
This has the effect of increasing the nucleation barrier for a second vortex. In fact, in a clean sample, the entry barriers for the first and the second vortex are nearly the same. 
However, in the presence of inhomogeneities, we can calculate how the first vortex is pinned near the surface with the effect of increasing the nucleation barrier for the second one, as shown in Fig. \ref{fig:nucleationBarriers}.
Hence, with this method, it is possible to accurately predict and design impurity density profiles able to protect the sample from vortex entry by pinning vortices in an area near the surface.

%
%
Real samples in general have rough surfaces. It is empirically known that in the presence of rough surfaces, the vortex entry barrier is altered \cite{schmidt2013physics}. \modified{A rough surface represents a challenging problem
for the calculation of both superheating fields \cite{Pack2020} and barriers \cite{bass1996effect}}

\begin{figure}[!hbt]
    \centering
    \includegraphics[width=0.99\columnwidth]{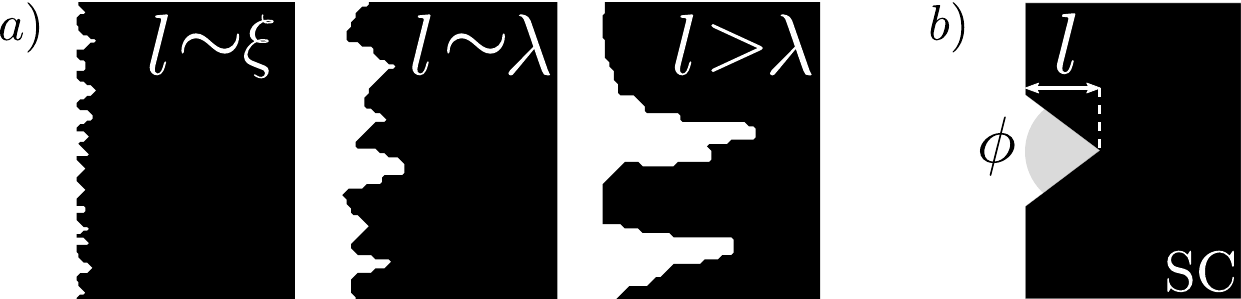}
    \caption{(a) Examples of roughness profiles at the superconductor's edge tested in the simulations. In these three cases the profile is, going from left to right, comparable to the coherence length $\xi$, comparable to the penetration depth $\lambda$, and bigger than $\lambda$.
    (b) Model of a single dent: The depth is parametrized by $l$, while its sharpness by the angle $\phi$. The black color indicates the superconductor (SC).}
    \label{fig:roughnessprofiles}
\end{figure}

For a quasi-two-dimensional superconductor, we can describe a rough edge as a sequence of geometrical defects, forming dents in the sample, as sketched in Fig. \ref{fig:roughnessprofiles}. The method is also straightforwardly applicable to the three-dimensional case.

The vortex entry occurs from a single dent, where the barrier is suppressed the most.
Therefore, it is relevant to study how the geometrical properties of one dent affect the nucleation barrier $\Delta F_n$. 
As shown in Fig. \ref{fig:roughnessprofiles}(b), we can characterize the dent by its depth $l$ which can be comparable to or bigger than the characteristic length scales of the model (coherence length $\xi$ and penetration depth $\lambda$), and by an angle $\phi$ which determines its sharpness.
\begin{figure}[!hbt]
    \centering
    \includegraphics[width=0.99\columnwidth]{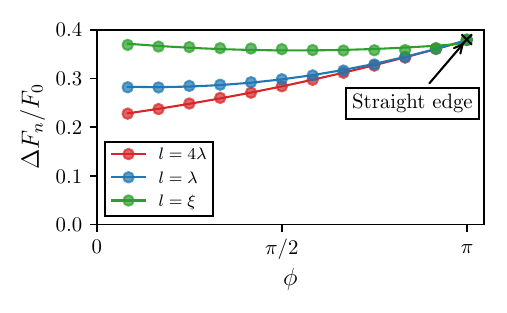} 
    \caption{Nucleation barrier computed for a dent with depth $l$ as a function of the angle $\phi$ as shown in Fig. \ref{fig:roughnessprofiles} (b). The system parameters are $\kappa=5$ and $H=0.4H_c$. The continuous lines are obtained by a quadratic regression (OLS) and agree with the input data. In particular, for the bigger dent, the nonlinear coefficient is negligible and the behavior is substantially linear.}
    \label{fig:roughness}
\end{figure}
Figure \ref{fig:roughness} compares the nucleation barriers $\Delta F_\textrm{n}/F_0$ for different dent depths $l$, as a function of the sharpness angle $\phi$.
For $\phi=\pi$ we have the limit of a perfectly straight edge.
The green line shows the nucleation barrier for $l=\xi$. We notice that the energy barrier is unaltered by the presence of the dent with a comparable or smaller roughness. 
Hence, roughness profiles with depths of the order of a coherence length have no effect on the nucleation barrier, independently of the sharpness of the dents.
As $l$ increases, the barrier suppression becomes substantial, depending on the sharpness $\phi$. For $l=\lambda$ the dependence on $\phi$ is non-linear, whereas for $l>\lambda$ (in our case $l=4\lambda$), the barrier decreases linearly with $\phi$.
%
In conclusion, we formulated a method to compute the minimum energy path in a gauge theory. We applied this method to solve the problem of vortex nucleation in the Ginzburg-Landau model of superconductivity. This is, a solution of the full nonlinear problem of the vortex entry barrier. 
The method allows calculating vortex entry barriers in the presence of surface roughness and impurities.
We showed that previously developed models to estimate the vortex entry barrier, based on the London approximation, are, in general, inadequate to describe the nucleation process even for a perfectly flat surface.  
We find that the surface roughness at the scale of coherence length affects the barriers insignificantly, and thus for superconductors with short-scale surface roughness, the main mechanism for the surface barrier reduction is the modulation of the superfluid density at larger length scales near the surface. The method straightforwardly applies
to three-dimensional configurations and to the geometries where demagnetization fields are important \cite{PhysRevLett.122.187702}. Such vortex entry barriers are of key importance in the design of quantum devices such as single-photon detectors and qubits, as well as superconducting rf cavities, transmission lines, and magnets.   

Finally, the method  is straightforwardly applicable to other gauge theories, different boundary conditions \cite{Samoilenka2020} which includes microscopic models of superconductivity, and sphaleron identification in high-energy physics.
\begin{acknowledgements}
We thank Mats Barkman, Robert Vedin, and Oskar Palm for useful comments. We gratefully acknowledge the support of NVIDIA Corporation with the donation of the Quadro P6000 GPU used for this research. Published with the support from the Längman Culture Foundation. The work was supported by the Swedish Research Council Grants No. 642-2013-7837, No. 2016-06122, No. 2018-03659, and G\"{o}ran Gustafsson Foundation for Research in Natural Sciences and Medicine and Olle Engkvists Stiftelse.
A.M. and A.B. contributed equally to the realization of this work.
\end{acknowledgements}

\section{Appendix}
\subsection{The Ginzburg-Landau Model}
The effective   free energy Ginzburg-Landau functional describing a superconductors reads:
\begin{align}
\label{eq:SI2-GLmodel}
F[\vb A, \Psi] =  d\int_\Omega & \dd x \dd y \bigg( \frac{1}{2m} \left| \left(-i\hbar\nabla + q \vb A\right) \Psi \right|^2  + \alpha \abs{\Psi}^2 \nonumber \\ 
& + \frac{\beta}{2} \abs{\Psi}^4 + \frac{(\curl \vb A - \mu_0 \vb H)^2}{2\mu_0}\bigg)\,.
\end{align}
The complex field $\psi$ is the superconducting order parameter, coupled to the electromagnetic field through the vector potential $\vb A$. Here $d$ is the effective thickness of the sample with a cross-section denoted by $\Omega$, $m$ is the mass of a superconductive carrier (\textit{i.e.}, the Cooper pair), $q$ is the coupling constant with the gauge field, $\alpha$ and $\beta$ are two temperature-dependent parameters, $\vb H$ is the uniform applied field while $\hbar$ and $\mu_0$ are, respectively, the Planck constant and the vacuum permeability. We consider the case $\vb H\parallel \hat{\boldsymbol{e}}_z$. 

To make the model dimensionless we scaled the variables as follows:
\begin{align}
F &= F_0 f, \nonumber \\
\mathbf{A} &= \mu_0  H_\textrm{c} \lambda \mathbf{a}, \nonumber \\
\Psi &= \psi_0 \psi, \nonumber
\end{align}
and the spatial coordinates become $x\rightarrow \lambda x$ and $y\rightarrow \lambda y$. $\psi_0 = \sqrt{\frac{-\alpha}{\beta}}$ is the uniform state order parameter, $\lambda = \frac{\sqrt{m}}{\sqrt{\mu_0}q\psi_0}$ is the penetration depth, $H_\textrm{c} = \sqrt{\frac{1}{\mu_0}\frac{\alpha^2}{\beta}}$ is the thermodynamic critical field, $ F_0 = \mu_0H_\textrm{c}^2\lambda^2 d$ is the characteristic energy value. 
We also introduce the Ginzburg-Landau parameter $\kappa = \frac{\lambda}{\xi}$ where $\xi = \frac{\hbar}{2\sqrt{-\alpha m}}$ is the coherence length. With this choice of definition  for the coherence length \cite{svistunov2015superfluid}, the critical coupling which separates type-I and type-II superconductors corresponds to $\kappa=1$. 

The resulting dimensionless Ginzburg-Landau model is expressed, up to constant terms, by the following free energy functional: 
\begin{equation}
\label{eq:GLmodel}
\begin{split}
f[\vb a, \psi] = \int_\Omega \dd{x} \dd{y} \Big(\frac{1}{2} \left| \left(-i\frac{2}{\kappa}\nabla + \vb a\right) \psi \right|^2 + \\ + \frac{1}{2} \qty( 1 - \abs{\psi}^2)^2 + \frac{(\curl \vb a - \vb h)^2}{2}\Big)\,,
\end{split}
\end{equation}
 where ${\vb h} = {\vb H}/H_\mathrm{c}$.
Quantities  in  capital
letters are measured in SI units, while variables in lower
case are dimensionless.
The Ginzburg-Landau model features local $U(1)$ gauge symmetry. 
This means that $\vb a$ and the phase of $\psi$ do not represent physically observable quantities. Instead, the observable fields of the system are the magnetic field $\vb b = \curl \vb a$ and the superconductive current density $\vb j = - \frac{2}{\kappa} \Im{\psi^* \qty(-i \frac{2}{\kappa} \grad + \vb a) \psi}$, which are invariant to the local gauge transformation:
\begin{equation}\label{eq:gaugeTransformation}
\begin{split}
    \psi^\prime   &= \textrm{e}^{i\frac{2}{\kappa}\chi}\psi\\
    \vb a ^\prime &= \vb a - \grad \chi\, ,
\end{split}
\end{equation}

In a type-II superconductor, when $H < H_\mathrm{c1}$ the equilibrium state is in the Meissner phase, no vortices are present in the bulk, and thus the magnetic field is completely screened by the superconducting current at surface. For $H > H_\mathrm{c1}$, the system is in the Shubnikov phase in which vortices are present. However, due to the presence of the surface barrier the Meissner state can survive in a meta-stable way for higher fields than $H_\textrm{c1}$, i.e. up to the \emph{spontaneous nucleation field} $H_\textrm{n}$. In the absence of fluctuations, only when this field is exceeded the barrier is suppressed and vortices can nucleate freely from the boundaries. Moreover, if the applied external field is higher than the \emph{spontaneous escape field} $H_\textrm{e}$ vortices cannot spontaneously escape from the sample without overcoming the surface barrier. This means that in the region $H_\textrm{e}<H<H_\textrm{n}$ Shubnikov states are metastable as the presence of the surface energy barrier prevents changes in the number of vortices.

\subsection{Minimum Energy Paths}
Vortex nucleation in a magnetically superheated superconductor is an example of metastable state decay.  
Various approaches to the study of metastable decay have been proposed \cite{kramers1940brownian, langer1969statistical}, and they identify the transition rate as $\tau^{-1} = A \exp(-\frac{\Delta E}{k_\mathrm{B}T})$, where $A$ is a prefactor.
$\Delta E$ is the energy barrier that has to be crossed for the transition to occur. In a system whose phase space is multidimensional, the barrier corresponds to a saddle point of the energy landscape, while in field theory, the infinite-dimensional saddle point is called sphaleron. Since the energy barrier appears as the argument of the exponential, the first step in the study of rare events is the sphaleron identification. 
A way to achieve this is the computation of the \emph{minimum energy path} (MEP) between two states. The MEP is a path in the configuration space such that it crosses the minimum in the cotangent space of the path point by point. 
Considering a system described by a potential $f(\vb q)$ where $\vb q$ is the configuration of the system, the mathematical definition can be written like
\begin{equation}
    \vb q(s) = \arg \min f_\bot [\vb q(s)]\,, \quad \forall s\, \in [0,1]\,,
\end{equation}
where the \emph{transition coordinate} $s$ has been introduced. The function $f_\bot$ is the potential function restricted to the cotangent space of the trajectory in point $\vb q(s)$. 

\subsection{Simplified string method}

The most common family of methods to compute minimum energy paths are the so-called \emph{chain-of-states} methods. These methods are based on the constrained optimization of a collection of systems $\{\vb q_n\}$, called \textit{frames}, subjected to the same potential $V(\vb q)$ but a different fictitious force. In this way, it is possible to capture aspects of dynamics of a transition without involving real-time dynamics~\cite{koslover2007comparison, vanden2010transition}. 
These methods can be summarized as variants of the widely known Nudged Elastic Band method \cite{sheppard2008optimization,BESSARAB2015335}. 
However, nudged elastic band techniques require an additional undefined constant (string elasticity) for introducing fictitious forces. Besides, the chain can miss the saddle point and trying to prevent this requires additional considerations  \cite{doi:10.1063/1.1329672}. 

The approach we use is the \textit{simplified string method}  \cite{weinan2007simplified}, generalized to a gauge field theory, in which we evolve the entire path $\vb q (s)$ in the configuration space towards the minimal energy one. The variable $s$ continuously parametrizes the path which connects the initial state $\vb q (0)$ to the final state $\vb q (1)$. 
It is crucial to notice that, when talking of evolution, we refer to a \textit{pseudo-dynamics} of the system and not real-time dynamics. 
To evolve the initial path to the minimal energy one, we apply a gradient descent algorithm according to:
\begin{equation}\label{eq:descent}
\pdv{\vb q}{\tau} = - \grad{f} + \lambda(\vb q, s) \pdv{\vb q}{s}\,.     
\end{equation}
Here $\tau$ is the pseudo-time describing the evolution of the string towards the MEP. The quantity $f[\vb a, \psi]$ is the free energy defined in Eq. \eqref{eq:GLmodel}, $\vb q(s)=\qty(\vb a(\vb r,s), \psi(\vb r, s))$ and $\grad{f}=\qty(\fdv{f}{\vb a},\fdv{f}{\psi^*})$. 
The term $ \lambda(\vb q, s) \pdv*{\vb q}{s}$ has the function of maintaining a certain parametrization of the path $\vb q(s)$. Since $\pdv*{\vb q}{s}$ is uniquely directed along the tangent direction, $\lambda$  does not affect the trajectory of the string. 

\subsubsection{Algorithm}
In the numerical implementation the path is discretized in a collection $\{\vb q_n \}$ of $N$ frames such that $\vb q_n \equiv \vb q \left(\frac{n}{N-1}\right)$. At each iteration, the algorithm performs a minimization procedure and a reparametrization. In the minimization phase, a certain number of gradient descend steps are performed. In the reparametrization phase, the displaced frames are used to compute a new path by an interpolation method. Usually, the cubic spline interpolation is employed as it guarantees a smooth curve. The interpolation function is then used to generate a new collection of frames that are chosen to be equidistant with respect to the selected metric. There is a certain freedom in the choice of the metric, which we analyze shortly.
To describe more in detail how the string method works, let us define $\vb q^{i,k}_n$ as the frame number $n$ at iteration $i$, while $k$ is the index of minimization steps. The simplified string method iteration works in the following way:
\begin{enumerate}
    \item \textbf{Minimization phase} \\
    Perform $M$ minimization steps using a gradient descent method:
    \begin{equation}
        \vb q_n^{i,k+1} = \vb q_n^{i,k} + \alpha_n^{i,k} \vb d_n^{i,k}
    \end{equation}
    where $\vb d_n^{i,k}$ is the descent direction and $\alpha_n^{i,k}$ is the step length. At the end of the minimization step the displaced frames $\tilde{\vb{q}}_n^{i}=\vb{q}^{i,M}_n$ have been computed.
    
    \item \textbf{Reparametrization step} \\
    Let $\norm{\cdot}$ be the norm induced by a chosen metric, the displacements between the frames are defined as $\Delta s_n = \norm{\tilde{\vb q}_n^i -\tilde{\vb q}^i_{n-1}}$. The parametrization is then computed as
    \begin{equation}
        s_n = \frac{\sum_p^n \Delta \tilde{s}_p}{\sum_j^F \Delta \tilde{s}_j}\,.
    \end{equation}
    
    In this way the curve is parametrized as $\vb q_n \equiv \vb q(s_n)$ with $s\in[0,1]$. Using the input set $I=\{(s_n, \vb q_n)\}$ the interpolating function $\hat{\vb q}^i(s)$ is estimated. Finally, a new set of equidistant frames is generated by
    \begin{equation}
        \vb q_n^{i+1} = \hat{\vb q}^i\left(\frac{n}{F-1}\right)\,.
    \end{equation}
\end{enumerate}
%
%
\subsubsection{Metric and Gauge Invariance}
The application of the string method to sphalerons identification in gauge theories is associated with several technical challenges, mainly related to metric definition for a Hilbert space and gauge invariance preservation. For this reason, we have developed a variant that we called \emph{gauged string method}, which we used to compute physically meaningful minimum energy paths while avoiding the use of global gauge fixing. If we consider a generic path in the phase space $\vb q(s)$, the system may evolve in $s$ changing the gauge of the solution without affecting the observable quantities of the system. In other words, the string gets twisted in the gauge degree of freedom crossing zones of the configuration space where there is no movement in the physical fields.  

This can be better understood thinking of the role of the metric, necessary to calculate the distances $\Delta s_n = \norm{\tilde{\vb q}_n^i -\tilde{\vb q}^i_{n-1}}$ needed for the reparametrization step. In a straightforward extension of the prescription of the simplified string method to a Hilbert space, the metric for the Ginzburg-Landau free energy is the distance defined as: 
\begin{equation}\label{eq:gaugeDependent}
    \Delta s^{gd}_n =   \qty(\norm{\vb a_n - \vb a_{n-1}}^2+\norm{\psi_n - \psi_{n-1}}^2)^{1/2}\,,
\end{equation}
where $\norm{\cdot}$ is the standard $L^2$ norm. The problem arising with this choice is the use of the gauge dependent fields $\vb a$ and $\psi$. This in a way means that two configurations of the same physical system expressed in two different gauges are considered as two distinct systems. Hence, the string can move in the gauge degree of freedom.

\begin{figure}[!hbt]
    \centering
    \includegraphics[width=1\columnwidth]{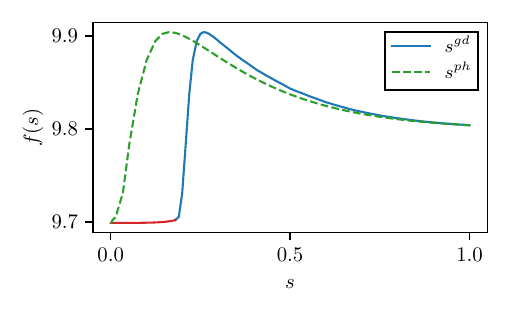} 
    \caption{The same minimum free energy path $f(s)$ is plotted with the gauge dependent metric $s^{gd}$ defined in Eq. \eqref{eq:invatianMetric1} and gauge independent metric $s^{ph}$ defined in Eq.\eqref{eq:ourmetric}. The red highlighted zone at the beginning of the $f(s^{gd})$ path collapses to a single point in the $f(s^{ph})$. This means that the pseudo-motion in that zone is an artefact due to the gauge symmetry. Indeed, in the red zone the system moves along the nonphysical gauge degree of freedom.}
    \label{fig:example_gt}
\end{figure}

There can be a region of the string where the fields $\vb a$ and $\psi$ are changing, but there is no variation in the observable quantities, such as $\vb b$  and $\vb j$.
The approach we use is to define a new gauge-invariant metric. With respect to the gauge transformation in Eq. \eqref{eq:gaugeTransformation} 
a possible choice, based on Eq.\eqref{eq:gaugeDependent}, reads:
\begin{align}\label{eq:invatianMetric1}
\begin{split}
    \Delta s^{ph}_n &= \Big(\norm{\abs{\psi}_n - \abs{\psi}_{n-1}}^2\\
                                    &+\norm{\qty(\frac{2}{\kappa}\grad{\theta_n} - \vb{a}_n)  -\qty(\frac{2}{\kappa}\grad{\theta_{n-1}} - \vb{a}_{n-1}) }^2 \Big)^{1/2}\,.
\end{split}
\end{align}
Eq. \eqref{eq:invatianMetric1} entails some limitations. In fact, let us suppose to be interested in studying samples with particular geometry. Inside holes, or other domain exclusion, the order parameter $\psi$ is not defined. Hence, Eq. \eqref{eq:invatianMetric1} becomes $\Delta s^{ph}_n =  \qty(\norm{\vb a_n - \vb a_{n-1}}^2)^{1/2}$, which is clearly not gauge invariant. Therefore, we can modify the definition of the metric in Eq. \eqref{eq:invatianMetric1}, to obtain a gauge invariant quantity which is related to the order parameter $\psi$ and simultaneously defined both in the domain and geometrical exclusions. By using the observables $\vb b$ and $\vb j$ we define:
\begin{equation}\label{eq:ourmetric}
    \Delta s^{ph}_n = \qty(\norm{\vb b_n - \vb b_{n-1}}^2+\norm{\vb j_n - \vb j_{n-1}}^2)^{1/2}\,.
\end{equation}
In the development of our method, we compared the outcomes of choosing different metrics. The sphaleron does not depend on it, as in Fig. \ref{fig:example_gt} displays. In fact, the height of the barrier is the same for $s^{gd}$ (continuous line) and $s^{ph}$ (dashed line), with the definition of Eq. \eqref{eq:ourmetric}, and it remains the same also for the metric in Eq. \eqref{eq:invatianMetric1}.
If we focus on the line obtained using the metric $s^{gd}$, we notice that the first part, highlighted in red, is an artifact that corresponds to pure gauge evolution. 
%
%
\subsection{Details of the numerical method developed}
The energy landscape of the GL model is a system far more complex than the usual molecular dynamics problems addressed by the string method. The problem for 2D systems involves four different fields, it is strongly non-linear, and has a local gauge symmetry. For this reason, we need to introduce some expedients to apply the string method to this problem. One is the gauge-independent metric described above, and the others concern the numerical implementation of the algorithm.

In our computation, the string $\vb q(s)$ is discretized as a vector of $F$ frames $\vb q_n$. The number of frames we employ ranges between 80 and 120, depending on the transition we are considering. Each of these frames, $\vb q_n = \qty(a_x, a_y, \Re{\psi}, \Im{\psi})$ represents a state of the system and is discretized on a mesh grid with a minimum of $400\times400$ lattice sites. We adopted a second-order finite difference scheme for the discretization of the functional. Then the gradient force is computed by simple analytical derivation of the function obtained by the discretization process.

For the minimization step, we applied the nonlinear conjugate gradient (NLCG) method with the Polak-Ribière-Polyak condition with automatic reset and exact line search. NLCG is necessary as the steepest gradient descent is too slow for this type of problem. Notice that, since the frames are displaced in the reparametrization step, the conjugacy is lost, and the conjugate direction needs to be reset at each iteration. In the final steps of the simulation, conjugacy can be switched off. By doing this, the algorithm applies the steepest gradient as descent direction. 
For what concerns the reparametrization step, we decided to use a linear interpolation step. Linear interpolation is computationally cheaper than the usual cubic interpolation, used within the string method, and more numerically stable. The only negative side of this choice is that the continuity up to the second derivative of $\vb q(s)$ is lost. However, the loss of analytical smoothness is counterbalanced by the use of an high number of frames, which results in a more dense discretization along $s$, generating a regular and well-behaved MEP.

The initial guess we used the ansatz for a single  vortex $\psi(\vb r) = \tanh(- (\vb r - \vb r_0)^2) (x - x_0 + \textrm{i}(y-y_0))/|\vb r - \vb r_0|$ where $\vb r_0$ is the center of the vortex and it varies frame by frame. It starts from a point outside the superconductor at $s=0$ and moves toward the center of the specimen for $s=1$.
Concerning the numerical extrema of the Hamiltonian~(\ref{eq:GLmodel}), the current density perpendicular to the boundary may slightly deviate from zero due to the limited accuracy of the finite-difference scheme. 
This discrepancy tends to zero with increasing mesh density. 
Strictly speaking, for non-extremum cases, i.e., configurations along the MEP except initial,  sphaleron (saddle point) and final states, this behavior is not guaranteed. 
Therefore, we monitored the perpendicular component of the current, and it turned out that this value is negligible.
The computer implementation is developed using the General Purpose GPU (GPGPU) paradigm. In particular, the core of the tool is implemented in CUDA C++ language and run on a NVIDIA Quadro P6000 GPU.
%
%
\subsection{Additional results}

\subsubsection{Quantitative analysis of the planar edge problem}
We simulated a 2D square system with side length $L=24\lambda$. In such a vast domain, the interactions between the entering vortex and the other edges of the sample are negligible, allowing the study of the surface barrier of a flat edge, which is the standard case studied in the previous literature.

The minimum free energy path of the vortex nucleation process, in this case, presents only one sphaleron whose free energy is $F_\textrm{sphaleron} = \max_s F(s)$.
Once we identify the sphaleron, we compute the nucleation and escape barriers according to:
\begin{align}
    &\Delta F_\textrm{n} = F_\textrm{sphaleron} - F_\textrm{Meissner} \\
    &\Delta F_\textrm{e} = F_\textrm{sphaleron} - F_\textrm{Shubnikov}
\end{align}
where $F_\textrm{Meissner} = F(s=0)$ is the energy of the Meissner state while $F_\textrm{Shubnikov} = F(s=1)$ is the energy of the system with a vortex in the bulk. The nucleation barrier can easily be interpreted as the energy cost of inserting a vortex inside the domain while the escape barrier is the activation energy of the opposite transition. 
\begin{figure}[!hbt]
    \centering
    \includegraphics[width=0.9\columnwidth]{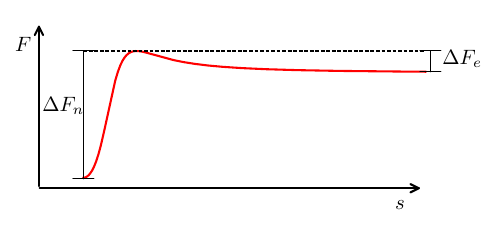}
    \caption{Sketch of a Minimum Free Energy Path for vortex nucleation with the definition of nucleation barrier $\Delta F_\textrm{n}$ and the escape barrier $\Delta F_\textrm{e}$.}
    \label{fig:deltas}
\end{figure}

Once these two quantities are computed for a wide region of the $\kappa$--$H$ space, it is possible to study the stability limits of the Shubnikov and Meissner phases, defining the spontaneous nucleation and escape fields as
\begin{align}
    \Delta F_\textrm{n}(H_\textrm{n}) = 0\,,\\
    \Delta F_\textrm{e}(H_\textrm{e}) = 0\,.
\end{align}
Externally applying $H_\textrm{n}$ generates a complete barrier suppression, and allows vortex entery from the boundaries, filling the sample until inter-vortex repulsion force prevents further nucleations. When $H_\textrm{e}$ is applied to the superconductor, solutions with vortices are not stable anymore, and they spontaneously escape from the bulk. Notice that, in principle, $H_\textrm{n}$ and the superheating field $H_\textrm{sh}$ are different quantities. The former is computed studying the entry process of the first vortex, while the latter is the mathematical stability limit of the Meissner state. These two quantities are strictly the same only for type-II superconductors.

\begin{figure}[!hbt]
    \centering
    \includegraphics[width=1.0\columnwidth]{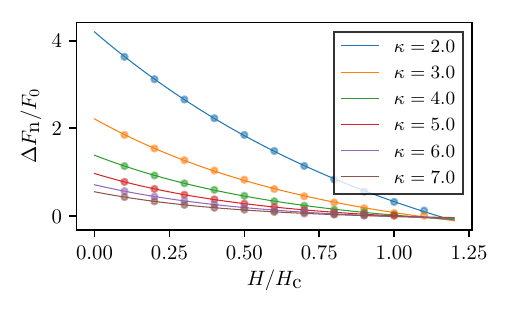}
    \caption{Nucleation barrier $\Delta F_\textrm{n}$ for various GL paramters $\kappa$ as a function of the external field $H$. The continuous lines are the estimated exponential regressions with ansatz $\Delta F^{\textrm{(fit)}}_\textrm{n}(H)/F_0 = C_1 \exp\qty(-C_2(H-H_\textrm{n}))$ where $C_1$, $C_2$ and $H_\textrm{n}$ are the coefficients estimated for each value of $\kappa$. Here $H_\textrm{c} = \sqrt{\frac{1}{\mu_0}\frac{\alpha^2}{\beta}}$ is the thermodynamic critical field while $F_0 = \mu_0H_\textrm{c}^2\lambda^2 d$ is the characteristic energy value.}
    \label{fig:deltaFn-kH-all}
\end{figure}

We have simulated this transition in a wide region of the $\kappa$--$H$ region, and show the results in Fig. \ref{fig:deltaFn-kH-all}.

From the computed $\Delta F_\textrm{n}$ is possible to extrapolate the nucleation field $H_n$. To do that, we interpolate the data points with the ansatz $\Delta F^{\textrm{(fit)}}_\textrm{n}(H)/F_0 = C_1 \exp\qty(-C_2(H-H_\textrm{n}))$ where $C_1$, $C_2$ and $H_\textrm{n}$ are the coefficients to be estimated for each value of $\kappa$. The calculated nucleation field is within a 2\% error compared the superheating field computed in \cite{transtrum2011superheating}.

\begin{figure}[!hbt]
    \centering
    \includegraphics[width=1.0\columnwidth]{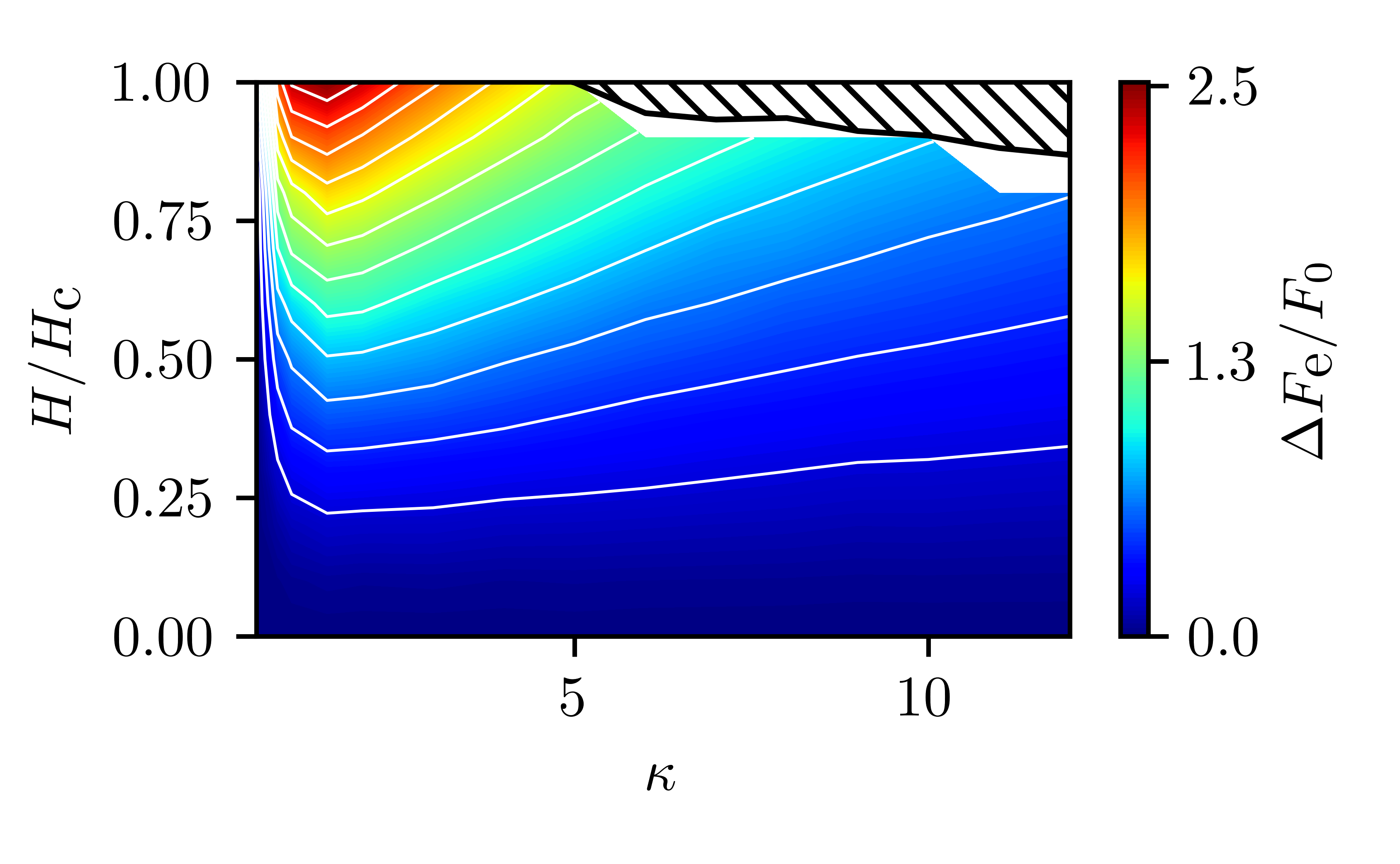}
    \caption{
    Escape barrier $\Delta F_\textrm{e} = \Delta F_\textrm{e}/F_0$ as a function of the external field $H$ and GL parameter $\kappa$. The space of parameters in the hatched zone in the upper right is unstable as $H>H_\textrm{n}$ and therefore a single vortex state is not a stable solution as there is no barrier preventing additional vortices to enter in the bulk. Notice that the contour-plot is cut just below for numerical reasons.}
    \label{fig:deltaFe-kH}
\end{figure}

We defined the escape barrier $\Delta F_\textrm{e} = F_{\text{sphaleron}}-F_{\text{Shubnikov}}$ as the barrier crossed in the event of the vortex exit from the bulk. The results for the barrier height in the $\kappa-H$ diagram are shown in Fig. \ref{fig:deltaFe-kH}.

Studying the escape barrier as a function of the external field $H$, we can notice as the barrier is exactly zero only for null external fields. This is in agreement with experimental results showing that weak external magnetic fields are sufficient to keep a vortex inside a superconductor. This result fixes the escape field $H_\textrm{e}=0$ meaning that an applied field in the opposite direction is needed to expel a vortex from a sample.

\begin{figure}
    \centering
    \includegraphics[width=\columnwidth]{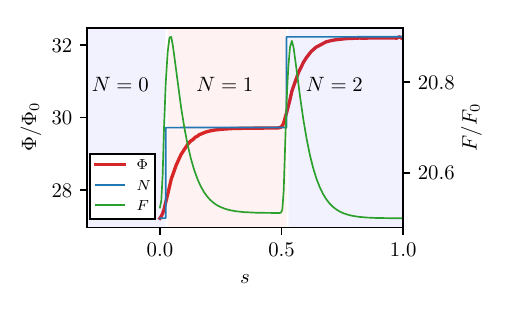}
    \caption{Evolution of the total magnetic flux $\Phi(s)$ across the sample along the minimum energy path during the process of nucleation of two vortices (red line). The free energy $F(s)$ (blue line) and the winding number $N(s)$ (green line) are also included.
    The flux is expressed in units of $\Phi_0 = \mu_0 H_c \lambda^2$.
    The sphaleron corresponds to to vortex carrying around 40\% of the flux quantum $\Delta \Phi_\textrm{q} = \kappa/2 \Phi_0$.
    The parameters of the system  are $\kappa=5$, $H=0.4H_\textrm{c}$}
    \label{fig:flux-nonquantization}
\end{figure}

The free energy $F(s)$ is not the only interesting quantity which can be studied  with this method. For example, it is well known that in the nucleation process the quantization of the magnetic flux does not hold near a surface \cite{schmidt2013physics}, which can be understood via a mapping of the problem to a vortex-(image)antivortex  pair. The variation of the total magnetic flux $\Phi(s)$ along the Minimum Energy Path can be calculated as displayed in Figure \ref{fig:flux-nonquantization}. Comparing the curves $F(s)$, $N(s)$ and $\Phi(s)$, we notice that the nucleation mechanism involves a gradual increase  of the magnetic flux of the vortex.

%
%
\subsubsection{Effects of nonlinearity in vortex nucleation in the simplest case of semi-infinite system}

\begin{figure}[!hb]
    \centering
    \includegraphics[height=0.6\columnwidth]{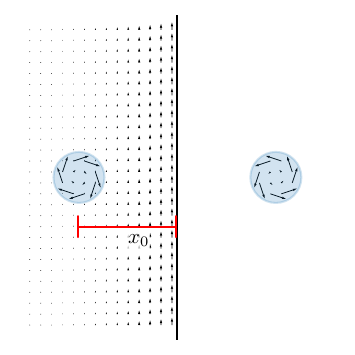}
    \caption{Sketch of the Bean-Livingston approach for the vortex nucleation process. In the London limit the free energy is obtained considering the interaction of the vortex with the combination of the decaying surface current and the image-antivortex. The image-antivortex is placed symmetrically with respect to the surface and its presence is necessary to impose the correct boundary conditions.} 
    \label{fig:bl-model}
\end{figure}

To assess the effects of nonlinearity, we consider the nucleation problem for the planar surface for which the  linear model from Bean and Livingston was constructed \cite{bean1964surface}. 
The Bean-Livingston model adopts an image-charge method. In this approach the energy of the vortex can be seen as the sum of the interaction of the vortex at position $x_0$ with the decaying external field and the interaction with an image anti-vortex at position $-x_0$ as shown in \ref{fig:bl-model}.

We can write the free energy as a function of the position in SI units as 
\begin{equation}
\begin{split}
F(x_0) \simeq - \frac{1}{2} \frac{1}{2\pi} \frac{\Phi_0^2}{\lambda^2} K_0 \left(\frac{2 x_0 }{\lambda}\right) + \\ + \Phi_0 H e^{-\frac{x_0}{\lambda}} + H_{c1} \Phi_0 - H \Phi_0\,.
\end{split}
\end{equation}
The first term is the interaction with the image-antivortex (with a $\frac{1}{2}$  factor needed to avoid double counting), the second is the interaction of the vortex with the decaying magnetic field at the surface, the third term is the energy of the magnetic field of the vortex while the last term is the interaction with the external field.

\begin{figure}[!hb]
    \centering
    \includegraphics[width=1.0\columnwidth]{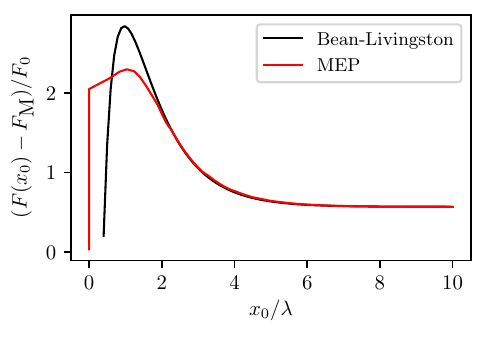}
    \caption{Free energy of a vortex as a function of the distance $x_0$ to the edge of the sample compared to the Meissner state free energy $F_\textrm{M}$. The black line is calculated using the Bean-Livingston model, while the red one is obtained by extrapolating the position of the vortex core from the minimum energy path. 
    An offset has been added to have the two curves coinciding for $x_0$ deep in the bulk, since we are comparing a London model result with Ginzburg-Landau results.
    The value matches for long distances while near the boundary, nonlinear effects  significantly diminish the value of the barrier. Moreover, the Bean-Livinston approximation fails to describe the steps of the process preceding the nucleation of a distinguishable vortex. This can be appreciated in the graph as the abrupt change from zero to about $2F_0$. The parameters of the simulation are $\kappa=1.3$ and $H=0.80 H_c$.
    }
    \label{fig:bl}
\end{figure}

To compare the Bean-Livingston picture with our results, we needed to extract a similar information from the computed minimum energy paths. This can be done by extrapolating, for each point of the minimum energy path, the position $x_0$ of the zero of the order parameter, which corresponds to the center of the vortex. In this way, we estimated the energy as a function of the position $F(x_0)$. 
Even for this simplest geometry, the linear models is inaccurate when there is a non-negligible interaction between the vortex core and the edge. For low-$\kappa$ the Bean-Livingston model fails to capture the mechanism of vortex nucleation as the vortex deformation near the boundaries plays a significant role in substantially decreasing the value of the nucleation barrier as can be observed in Fig. \ref{fig:bl}. Moreover, differently from the Bean-Livingston approach, The full Ginzburg-Landau model is able to describe also the transition states preceding the formation of a distinguishable vortex structure.
%
%

\subsubsection{Vortex nucleation in a L-shape geometry}

\begin{figure}[hbt]
    \includegraphics[width=\columnwidth]{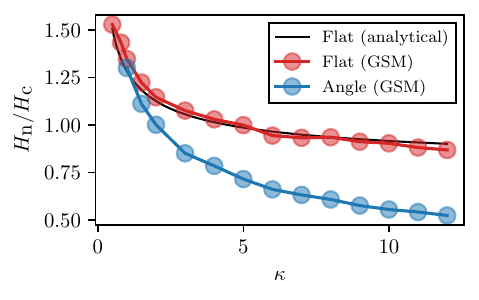} 
    \caption{
    Nucleation field computed by exponential fitting of the results of gauged string method (GSM) in the case of nucleation from a flat boundary and a L-shaped boundary as described in Figure \ref{fig:L-shape}. The analytical estimate by \cite{transtrum2011superheating} is plotted as reference. The gauged string method provide a reliable way to compute nucleation fields in arbitrary geometry. This quantity is accessible with simple experiments. 
    }
    \label{fig:hn}
\end{figure}

\begin{figure*}[hbt]
    \includegraphics[width=1.8\columnwidth]{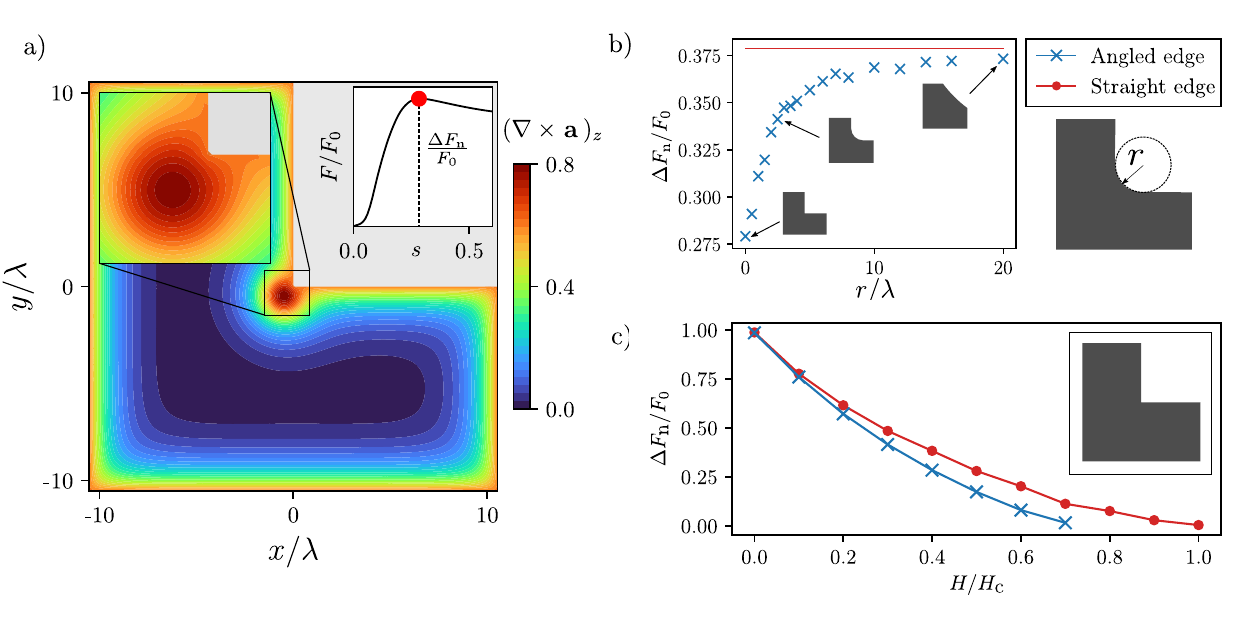} 
    \caption{
    Sphaleron and energy barriers in an L-shaped superconductor
    and in an L-shaped superconductor with a smoothed concave corner. 
    The concave angle is the most optimal entry point for the vortex. a) The magnetic field at the sphaleron.  b) The energy barrier as a function of the curvature radius $r$ of the notch, calculated at constant external magnetic field $H=0.4 H_\mathrm{c}$. c) The energy barrier for a superconductor with a straight edge in comparison with the case of concave-angled edge. We find that for the external magnetic field $H>0$, the barrier is always lower for the concave-angled geometry.
    Figure a) is obtained for Ginzburg-Landau parameter $\kappa = 3$, while figure b) and c) for $\kappa=5$. }
    
    \label{fig:L-shape}
\end{figure*}

A case of great interest for superconducting devices is a domain shaped as a L. In this case, as showed in Figure~\ref{fig:L-shape}\, which displays the sphaleron magnetic field configuration for vortex entry, the tip of the angle acts as nucleation center for the entering vortex. The L-shaped geometry is a well-studied benchmark for the numerical finding of critical fields \cite{GAO2015329}. The problem of sphaleron and energy barrier for this case was not solved.
Differently from the flat boundary case, in this situation, the vortex enters from the concave corner. To gain further insight into the dependence of the barrier on the sample geometry, we continuously deform the L-shaped geometry by varying the curvature radius of the concave corner, as shown in Figure \ref{fig:L-shape}\,b. For instance, curvature with a radius equal to zero corresponds to a sharp $\pi /2$ angle as in Figure \ref{fig:L-shape}\,a, while $r\gg 1$  eliminates concave corner and thus
should yield results similar to the flat-boundary situation. 
The vortex entry barrier strongly depends on the geometry of the sample for high external magnetic fields Figure~\ref{fig:L-shape}\,c. In general, for $H>0$, our numerical study shows that the vortex nucleation barrier increases with the curvature radius. By performing an exponential fit of the values of the nucleation barrier, it is possible to derive the nucleation field $H_n$ as shown in Figure \ref{fig:hn}. The nucleation fields is significantly lower with respect to the flat edge.

\bibliography{bibliography.bib}

\end{document}